\input{aipcheck}

\documentclass[
    ,final            % use final for the camera ready runs
  ]
  {aipproc}
\usepackage{pdflscape}
\usepackage{rotating}
\layoutstyle{6x9}
\def\hlf{{{1\over2}}}
\def\thlf{{{3\over2}}}

\begin{document}

\title{Symmetries of Doubly Heavy Baryons}


\author{Benjamin Eakins}{
  address={Department of Physics, Florida State University, Tallahassee, FL
32306}
}

\author{Winston Roberts}{
  address={Department of Physics, Florida State University, Tallahassee, FL
32306}
}

\begin{abstract}
For very heavy quarks, the two heavy quarks in a doubly heavy baryon are
expected to form a pointlike, heavy diquark in an antitriplet color
configuration. In this limit the dynamics of the light degrees of freedom
`factorize' from the dynamics of the heavy diquark system, and a superflavor
symmetry emerges which relates the properties of doubly heavy baryons to heavy
mesons. The charm quark may not be heavy enough for the $cc$ diquark in a
$\Xi_{cc}$ to be regarded as pointlike. However, there are indications from the
results of a nonrelativistic constituent quark model that many of the
consequences of factorization emerge even though the model does not assume a
quark-diquark structure and the mean separation of the two heavy quarks in the
model is not small. We discuss the consequences of factorization for the
spectroscopy of doubly heavy baryons and compare these consequences to results
from a quark model. We also discuss the possibility of treating the strange
quark as a heavy quark and applying these ideas to the $\Xi$.
\end{abstract}

\maketitle

%%%%%%%%%%%%%%%%%%%%%%%%%%%%%%%%%%%%%%%%%%%%
%% MAINMATTER
%%%%%%%%%%%%%%%%%%%%%%%%%%%%%%%%%%%%%%%%%%%%

\section{Introduction}
There are indications from models that a rich symmetry emerges for baryons with
two heavy quarks (Doubly Heavy Baryons or DHBs).  For infinitely heavy quarks
the two heavy quarks will form a pointlike heavy diquark \cite{SavageSpectrum}. 
If the heavy diquark is indeed pointlike then the physics of the light degrees
of freedom will be independent of the physics of the heavy diquark
\cite{eakinsroberts,kiselev02,gershtein00}.  Furthermore, one expects that the
total angular momentum of the heavy diquark will decouple from the total angular
momentum of the light degrees of freedom in analogy to Heavy Quark Symmetry
(HQS) \cite{eakinsroberts,IsgurWise}.  Therefore, one expects that the
properties of the light degrees of freedom will be independent of the diquark's
flavor, excitation state, and total angular momentum \cite{eakinsroberts}. 
There are indications from a model \cite{RobertsPervin} that a pointlike diquark
may not be a necessary condition for this Heavy Diquark Symmetry (HDS) to emerge
\cite{eakinsroberts,ERdecay}.

For this note we focus on the implications of this symmetry for pion emission. 
The decoupling of the total angular momentum of the heavy diquark implies that
states will exist in degenerate multiplets similar to those of HQS
\cite{eakinsroberts,IsgurWise,ERdecay}.  Partial widths of strong decay
transitions between multiplets will be related by spin-counting arguments
\cite{ERdecay}.  In practice it is possible to describe all of the pion-emission
decays between two multiplets in terms of a single amplitude.  These amplitudes
are expected to be independent of the heavy diquark's flavor and excitation
state.  We discuss this in more detail in Ref.~\cite{ERdecay}.  In this paper we
extract a few of these amplitudes using the wave functions from a
nonrelativistic quark model (the Roberts-Pervin or RP model
\cite{RobertsPervin}) and the $^3P_0$ model for transitions.

\newpage
%\begin{landscape}
\begin{table}[h]
\caption{Pion-emission HDS amplitudes for transitions from the parent multiplet
listed to the daughter multiplet listed.  The amplitudes are arranged in matrix
form with the parent baryon's position in the multiplet corresponding to the
column of the matrix and the daughter baryon's position in the multiplet
corresponding to the row of the matrix.  These decay amplitudes are extracted
from the $^3P_0$ model using quark model wave functions taken from the RP model
\cite{RobertsPervin}.  Phase space has been divided off as well as the
appropriate spin-counting factors. Thus, all of the amplitudes, in each section
of the table should be equal, according to lowest order HDS predictions.  The
entries in each set of parentheses are equal because of spin-counting arguments,
the entries in different columns are equal because of the diquark flavor
symmetry, and the entries involving different excitation states of the heavy
diquark are equal because of the excitation symmetry.  The decay modes omitted
are forbidden by spin-counting arguments.  In the section of the table below the
DHB results are analogous results for $\Xi$, $\Xi_{c}$, and $\Xi_{b}$ baryons
obtained by treating the strange quark as a heavy quark.  All of the amplitudes
given below correspond to a unique transition of the light degrees of freedom. 
Therefore, lowest order HDS predicts that they should all be equal.
\label{NoPStripwidths}}
{
\begin{tabular}{ccccccc}

\hline \hline
Parent Multiplet & $J_a^{P_a}$ & Daughter Multiplet & $\qquad J_b^{P_b} \qquad$
& \multicolumn{3}{c}{HDS Amplitude (GeV$^{-\hlf}$)} \\ 
\hline
 & & & & $\Xi_{cc}$ & $\Xi_{bc}$ & $\Xi_{bb}$ \vspace{1mm} \\ \cline{5-7} 

$1^{3}S_1\otimes 1P_\frac{1}{2}$ & $(\hlf^-,\thlf^-)$           
& $1^{3}S_1\otimes 1S_\frac{1}{2}$ & $\left( \begin{array} {c} \hlf^+ \\ \thlf^+
\end{array} \right)$
& $\left( \begin{array} {c} 0.708 \;\;  \quad-\quad  \\  \quad-\quad  \;\; 0.786
\end{array} \right)$  
& $\left( \begin{array} {c} 0.724 \;\;  \quad-\quad  \\  \quad-\quad  \;\; 0.738
\end{array} \right)$  
& $\left( \begin{array} {c} 0.748 \;\;  \quad-\quad  \\  \quad-\quad  \;\; 0.774
\end{array} \right)$ \vspace{2mm} \\  

$1^{1}P_1\otimes 1P_\frac{1}{2}$              & $(\hlf^+,\thlf^+)$           
& $1^{1}P_1\otimes 1S_\frac{1}{2}$ & $\left( \begin{array} {c} \hlf^- \\ \thlf^-
\end{array} \right)$ 
& $\left( \begin{array} {c} 0.607 \;\;  \quad-\quad  \\  \quad-\quad  \;\; 0.710
\end{array} \right)$  
& $\left( \begin{array} {c} 0.605 \;\;  \quad-\quad  \\  \quad-\quad  \;\; 0.702
\end{array} \right)$  
& $\left( \begin{array} {c} 0.600 \;\;  \quad-\quad  \\  \quad-\quad  \;\; 0.689
\end{array} \right)$ \vspace{2mm} \\ 

\hline

 & & & & $\Xi$ & $\Xi_{c}$ & $\Xi_{b}$ \vspace{1mm} \\ \cline{5-7} 

$1^{3}S_1\otimes 1P_\frac{1}{2}$ & $(\hlf^-,\thlf^-)$           
& $1^{3}S_1\otimes 1S_\frac{1}{2}$ & $\left( \begin{array} {c} \hlf^+ \\ \thlf^+
\end{array} \right)$
& $\left( \begin{array} {c} 0.486 \;\;  \quad-\quad  \\  \quad-\quad  \;\; 0.667
\end{array} \right)$  
& $\left( \begin{array} {c} 0.628 \;\;  \quad-\quad  \\  \quad-\quad  \;\; 0.653
\end{array} \right)$  
& $\left( \begin{array} {c} 0.637 \;\;  \quad-\quad  \\  \quad-\quad  \;\; 0.625
\end{array} \right)$ \vspace{2mm} \\ 

$1^{1}P_1\otimes 1P_\frac{1}{2}$              & $(\hlf^+,\thlf^+)$           
& $1^{1}P_1\otimes 1S_\frac{1}{2}$ & $\left( \begin{array} {c} \hlf^- \\ \thlf^-
\end{array} \right)$ 
& $\left( \begin{array} {c} 0.535 \;\;  \quad-\quad  \\  \quad-\quad  \;\; 0.635
\end{array} \right)$  
& $\left( \begin{array} {c} 0.564 \;\;  \quad-\quad  \\  \quad-\quad  \;\; 0.672
\end{array} \right)$  
& $\left( \begin{array} {c} 0.578 \;\;  \quad-\quad  \\  \quad-\quad  \;\; 0.678
\end{array} \right)$ \vspace{2mm} \\ 

\hline \hline
\end{tabular}
}
\end{table}
%\end{landscape}

\section{Discussion and Results}
Table \ref{NoPStripwidths} shows the pion-emission HDS amplitudes extracted from
the model for the case where the light degrees of freedom transition from an
excited state with angular momentum $J_l=\hlf^-$ to the ground state with
$J_l=\hlf^+$.  The top half of the table shows results for DHBs and the bottom
half shows analogous results obtained for $\Xi$, $\Xi_{c}$, and $\Xi_{b}$
baryons by treating the strange quark as a heavy quark.  Lowest order HDS
predicts that all of the amplitudes in the table below should be equal. 
However, the RP model does not assume a point-like diquark and it also includes
symmetry breaking spin interactions.  For DHBs (top half of Table
\ref{NoPStripwidths}) one can see that the effect of these symmetry breaking
interactions tends to decrease as the mass of the heavy quarks increase.  For
baryons in which the strange quark is treated as a heavy quark (bottom half of
Table \ref{NoPStripwidths}), spin dependent corrections seem to be much more
important as one would expect.

\subsection{Conclusions}
We have presented a small sample of HDS predictions.  In these results 24
partial widths are approximately determined by a single amplitude.  More results
are presented in Ref.~\cite{ERdecay}.

%%%%%%%%%%%%%%%%%%%%%%%%%%%%%%%%%%%%%%%%%%%%%%%%
%% BACKMATTER
%%%%%%%%%%%%%%%%%%%%%%%%%%%%%%%%%%%%%%%%%%%%%%%%

\begin{theacknowledgments}
We gratefully acknowledge the support of the Department of Physics, the College
of Arts and Sciences, and the Office of Research at Florida State University.
This research is supported by the U.S. Department of Energy under contract
DE-SC0002615.
\end{theacknowledgments}



\bibliographystyle{aipproc}   % if natbib is available

%\bibliography{sample}

% \IfFileExists{\jobname.bbl}{}
%  {\typeout{}
%   \typeout{******************************************}
%   \typeout{** Please run "bibtex \jobname" to optain}
%   \typeout{** the bibliography and then re-run LaTeX}
%   \typeout{** twice to fix the references!}
%   \typeout{******************************************}
%   \typeout{}
%  }
%%%%%%%%%%%%%%%%%%%%%%%%%%%%%%%%%%%%%%%%%%%%%%%%%%%%%%%%%%%%%%%%%%%
%%%%%%%%%%%%%%%%%%%%%%%%%%%%%%%%%%%%%%%%%%%%%%%%%%%%%%%%%%%%%%%%%%%
%%%%%%%%%%%%%%%%  MACRO FOR THE REFERENCES  %%%%%%%%%%%%%%%%%%%%%%%
%%%%%%%%%%%%%%%%%%%%%%%%%%%%%%%%%%%%%%%%%%%%%%%%%%%%%%%%%%%%%%%%%%%
%%%%%%%%%%%%%%%%%%%%%%%%%%%%%%%%%%%%%%%%%%%%%%%%%%%%%%%%%%%%%%%%%%%
\newif\ifmultiplepapers
\def\beginpapers{\multiplepaperstrue}
\def\endpapers{\multiplepapersfalse}  
\def\journal#1&#2(#3)#4{\rm #1~{\bf #2}\unskip, \rm  #4 (19#3)}
\def\trjrnl#1&#2(#3)#4{\rm #1~{\bf #2}\unskip, \rm #4 (19#3)}
\def\baps{\journal {Bull.} {Am.} {Phys.} {Soc.}&}
\def\jap{\journal J. {Appl.} {Phys.}&}
\def\prl{\journal {Phys.} {Rev.} {Lett.}&}
\def\pl{\journal {Phys.} {Lett.}&}
\def\pr{\journal {Phys.} {Rev.}&}
\def\np{\journal {Nucl.} {Phys.}&}
\def\rmp{\journal {Rev.} {Mod.} {Phys.}&}
\def\jmp{\journal J. {Math.} {Phys.}&}
\def\rmm{\journal {Revs.} {Mod.} {Math.}&}
\def\jetp{\journal {J.} {Exp.} {Theor.} {Phys.}&}
\def\sjetp{\trjrnl {Sov.} {Phys.} {JETP}&}
\def\dokl{\journal {Dokl.} {Akad.} Nauk USSR&}
\def\spd{\trjrnl {Sov.} {Phys.} {Dokl.}&}
\def\tmf{\journal {Theor.} {Mat.} {Fiz.}&}
\def\snp{\trjrnl {Sov.} J. {Nucl.} {Phys.}&}
\def\hpa{\journal {Helv.} {Phys.} Acta&}
\def\yf{\journal {Yad.} {Fiz.}&}
\def\zp{\journal Z. {Phys.}&}
\def\anp{\journal {Adv.} {Nucl.} {Phys.}&}
\def\ap{\journal {Ann.} {Phys.}&}
\def\am{\journal {Ann.} {Math.}&}
\def\nc{\journal {Nuo.} {Cim.}&}
\def\etal{{\sl et al.}}
\def\pre{\journal {Phys.} {Rep.}&}
\def\pca{\journal Physica (Utrecht)&}
\def\prs{\journal {Proc.} R. {Soc.} London &}
\def\jcp{\journal J. {Comp.} {Phys.}&}
\def\pna{\journal {Proc.} {Nat.} {Acad.}&}
\def\jpg{\journal J. {Phys.} G (Nuclear Physics)&}
\def\fort{\journal {Fortsch.} {Phys.}&}
\def\jfa{\journal {J.} {Func.} {Anal.}&}
\def\cmp{\journal {Comm.} {Math.} {Phys.}&}
%%%%%%%%%%%%%%%%%%%%%%%%%%%%%%%%%%%%%%%%%%%%%%%%%%%%%%%%%%%%%%%%%%%

\end{document}